# The living state: how cellular excitability is controlled by the thermodynamic state of the membrane

Christian FILLAFER[a,*], Anne PAEGER[a], Matthias F. SCHNEIDER[a]

[a]Medical and Biological Physics
Physics
Technical University Dortmund
Otto-Hahn-Str. 4
44227 Dortmund
Germany

**Address correspondence to:** Christian Fillafer; Physics - Medical and Biological Physics Technical University Dortmund; Otto-Hahn-Str. 4; 44227 Dortmund (Germany); tel: +49-231-755-2990; email: christian.fillafer@tu-dortmund.de

# Abstract

The thermodynamic (TD) properties of biological membranes play a central role for living systems. It has been suggested, for instance, that nonlinear pulses such as action potentials (APs) can only exist if the membrane state is in vicinity of a TD transition.

Herein, two membrane properties in living systems – excitability and velocity – are analyzed for a broad spectrum of conditions (temperature ($T$), 3D-pressure ($p$) and $pH$-dependence). Based on experimental data from Characean cells and a review of literature we predict parameter ranges in which a transition of the membrane is located ($15 - 35°C$ below growth temperature; $1 - 3\, pH$ units below $pH\, 7$; at $\sim 800\, atm$) and propose the corresponding phase diagrams. The latter explain: *(i)* changes of AP velocity with $T, p$ and $pH$. *(ii)* The existence and origin of *two qualitatively different* forms of loss of nonlinear excitability ("nerve block", anesthesia). *(iii)* The type and quantity of parameter changes that trigger APs. Finally, a quantitative comparison between the TD behavior of 2D-lipid model membranes with living systems is attempted. The typical shifts in transition temperature with $pH$ and $p$ of model membranes agree with values obtained from cell physiological measurements. Taken together, these results suggest that it is not specific molecules that control the excitability of living systems but rather the TD properties of the membrane interface. The approach as proposed herein can be extended to other quantities (membrane potential, calcium concentration, etc.) and makes falsifiable predictions, for example, that a transition exists within the specified parameter ranges in excitable cells.

# Abbreviations

AP – action potential
TD – thermodynamic
LE – liquid-expanded phase
LC – liquid-condensed phase

# 1. Introduction

This work deals with the central question: do cells have nonlinear excitability because their cell membrane can undergo a thermodynamic transition?

Membrane interfaces are generally responsive to environmental stimuli (mechanical, thermal, chemical, electrical, etc.). For some thermodynamic (TD) states of membranes, stimuli do not only lead to a linear response but – above a threshold – can result in all-or-none pulses (Mussel and Schneider, 2019a; Shrivastava and Schneider, 2014). Perturbations also propagate in cells and tissues. One of the most prominent nonlinear excitation phenomena is the action potential (AP). APs have been detected in different cells of animals (Aidley, 1998), plants (Wayne, 1994), protozoa (Schlaepfer and Wessel, 2015), etc. Remarkable experiments with internally perfused axons have demonstrated that APs can be excited despite removal of the bulk of intracellular material including the cortical cytoskeleton (Terakawa and Nakayama, 1985). This indicates that the propagation medium of the pulse is the cell membrane. Nonlinear pulses with the same characteristics as APs have been excited in phospholipid monolayers devoid of protein at the air-water interface (Shrivastava et al., 2015; Shrivastava and Schneider, 2014). The universal character of the phenomenon, *i.e.* the existence of nonlinear pulses in different cells and in simple model membranes indicates that nonlinear excitability *does not depend on specific molecules*. Therefore, it is of interest to ask which general properties a membrane must possess in order to support such pulses. Furthermore, it is of interest to identify the regime of states in which a membrane is excitable, in order to better understand the "living state" that has established itself in biological systems. It is the intent of the present work to approach this problem from a classical TD perspective following the work of Einstein (Einstein, 1910).

Theoretical predictions (Heimburg and Jackson, 2005; Kaufmann, 1989; Mussel and Schneider, 2019a, 2019b) and experimental evidence (Shrivastava et al., 2015; Shrivastava and Schneider, 2014) have indicated that nonlinear pulses can arise when a lipid membrane is perturbed close to an ordered-disordered transition. Heimburg and Jackson, in particular, proposed the existence of stable propagating solitary waves, when the membrane state is transiently forced into a transition regime (Heimburg and Jackson, 2005). Indeed, many synthetic phospholipid membranes, including those in which nonlinear pulses were observed (Shrivastava et al., 2015; Shrivastava and Schneider, 2014), exhibit transitions (*e.g.* (Lee, 1977)). There also exists evidence for transitions in biological membranes (platelets (Tablin et al., 1996), plant germ tips (Crowe et al., 1989), sperm cells (Crowe et al., 1989), bacteria (Heimburg and Jackson, 2005; Melchior and Steim, 1976), spinal cord (Music et al., 2019; Wang et al., 2018), etc.). Furthermore, calorimetry (Beljanski et al., 1997), optical studies (Georgescauld et al., 1979; Ueda et al., 1974) and membrane potential measurements (Inoue et al., 1973; Ueda et al., 1974) have indicated typical signs of transitions in excitable cells. However, the basis of evidence is too small, indirect and scattered for systematic conclusions. This is unfortunate, because it impedes central progress in our understanding of the origin of cellular excitability, non-excitable states of cells (anesthesia, etc.), means to stimulate cells, etc. We have proposed that a transition in an excitable membrane can be located from the "physiological phenomenology" (Fillafer and Schneider, 2013). The approach shall be explained briefly: in essence, we assume that physiological membrane functions (*e.g.* permeability (Mosgaard and Heimburg, 2013; Wunderlich et al., 2009), catalytic rate, excitability, pulse velocity) are determined by the TD state of the membrane interface. We call this "*the state-to-function relation*". Changes of the cellular conditions (*e.g.* by temperature ($T$), 3D-pressure ($p$), proton concentration ($pH$)) are expected to lead to more or less pronounced variations of the membrane state and thus functions. Particularly drastic changes will occur if the material undergoes a TD transition. The velocity of pulses, for example, should decrease upon entering a transition. The underlying reason is that the TD susceptibilities (*e.g.* compressibility, heat capacity), become maximal in the transition regime as compared to the neighboring phases (Albrecht et al., 1978; Steppich et al., 2010). At least in the linear case, the relation between pulse velocity and the isentropic compressibility $\kappa_S$ is

$$c \approx 1 / \sqrt{\rho \kappa_S} \qquad (1)$$

with $\rho$ as the density of the material. Therefore, if the velocity of pulses decreases nonlinearly, the system likely enters a transition regime. This correlation between compressibility of the system and pulse velocity has been confirmed for linear (Griesbauer et al., 2012) and interestingly also for nonlinear waves (Shrivastava and Schneider, 2014) in phospholipid monolayers.

It is the goal of the present work to analyze the thermodynamics of living, excitable systems in light of the relation between physical state and biological function (the *state-to-function* approach). In particular, we pursue the hypothesis that the origin of nonlinear cellular excitability is a transition of the cell membrane (a 2D-interface *consisting of lipids, proteins, carbohydrates, ions, water, etc*.). We first describe the theoretical concepts and predictions of the approach (3.1). Subsequently these predictions are tested qualitatively (3.2) as well as quantitatively (3.3-3.5) against our own experimental results from excitable plant cells (Characean internodes) and data from other authors. Two different mechanisms for loss of the nonlinear excitability of cells (anesthesia, nerve block) are proposed and the typical stimuli for cells are identified and interpreted.

## 2. Materials and methods

*Materials.* All reagents used were purchased from Sigma-Aldrich (St. Louis, MO, USA) and were of analytical purity ($\geq$ 99%).

***Cell cultivation and storage.*** *Chara australis* was cultivated in glass aquariums filled with a layer of soil ($\sim 2\ cm$), quartz sand ($\sim 1\ cm$) and deionized water. The plants were grown under illumination from an aquarium light (14 W, Flora Sun Max Plant Growth, Zoo Med Laboratories Inc., San Luis Obispo, CA, USA) at a 14:10 light:dark cycle at room temperature ($\sim 20\ °C$). Prior to use, single internodal cells were stored for a minimum of $12\ h$ in a solution containing $0.1\ mM$ NaCl, $0.1\ mM$ KCl and $0.1\ mM$ CaCl$_2$.

***Propagation velocity measurements.*** A setup similar to that described in (Lühring, 2006) was used. It consists of a compartmentalized plexiglass chamber which can be controllably heated/cooled and which can hold cells up to $10\ cm$ in length. Small extracellular sections (length $\sim 5\ mm$) of the cell were electrically isolated against each other with vacuum grease (Dow Corning Corporation, Midland, MI, USA). The K$^+$-anesthesia technique (Lühring, 2006) in combination with extracellular Ag/AgCl-wire electrodes was used for monitoring the membrane potential at two sites along the internode. The extracellular solutions contained $110\ mM$ KCl in the outmost compartment and artificial pond water in all other compartments (APW; $1\ mM$ KCl, $1\ mM$ CaCl$_2$, $190\ mM$ D-sorbitol, and $5\ mM$ buffer substance (MES for $pH\ 4.5-7$; TRIS for $pH\ 7-10$); pH was set to 7.0 with HCl/NaOH). The potential between the virtual intracellular electrode (KCl-compartment) and extracellular electrodes was recorded with voltage sensors (PS-2132; $100\ Hz$ sample rate; PASCO scientific, Roseville, CA, USA). When a supra-threshold electrical stimulus was applied, an AP was excited and subsequently passed along downstream electrodes. The delay time of the pulse between two electrodes with a known spatial separation is inversely proportional to the pulse propagation velocity. Since the conduction velocity varied between cells and since the main interest was on identifying relative changes, the data were normalized to the velocity in the basal state ($T = 20°C; pH = 7.0$).

To vary the temperature of the cell in a controlled manner, the measuring chamber was placed on a solid copper block which was perfused by a heat bath (RM6, Lauda GmbH, Lauda-Königshofen, Germany). The pH of the extracellular medium was varied by titration with HCl and NaOH respectively.

***Excitation by cooling.*** The medium in one of the compartments ($0.5 - 1\ mL$) was replaced with a micropipette by the same medium at a lower temperature. First, relatively large stepwise temperature drops were employed ($\Delta T \sim -5°C$). The criterion for excitation was the detection of an action potential (amplitude $\sim 100 - 150\ mV$) with a voltage sensor at a distance of $2 - 3\ cm$. Once a supra-threshold step was reached, attempts were made to confirm/narrow down the threshold by similar/smaller temperature drops. In between different trials, the cell was allowed to re-equilibrate at the basal temperature for $\sim 10\ min$. Typically, a cell was exposed to $\sim 15$ temperature steps.

## 3. Results and Discussion

### 3.1. Theory. TD state controls cellular excitability

Simply put, the physical theory of nerve pulse propagation – as opposed to the electrical circuit theory – rests on the idea that *energy-, entropy- and momentum conservation must not be violated in biology*. Since a membrane is decoupled from its surroundings (Griesbauer et al., 2009), a perturbation of the membrane will propagate. This follows straight from fundamental physical principles. Wilke was presumably one of the first to mention "adiabatic propagation" in the context of nerve pulses as early as 1912 (Wilke, 1912), but others like A.V. Hill commented along the same lines (Hill, 1912). The key theoretical step came by Kaufmann in 1989 who put the idea on a thorough theoretical footing (Kaufmann, 1989). From there, Heimburg & Jackson developed the "soliton model" of nerve pulse propagation (Heimburg and Jackson, 2005). Apart from some details (*e.g.* we do not believe that the propagating pulse is a soliton (Fillafer et al., 2017; Shrivastava et al., 2018)) we agree with their framework. The key idea is that APs are nonlinear sound pulses that propagate in the membrane. The origin of the nonlinearity of the response is a TD transition of the membrane interface (Heimburg and Jackson, 2005; Kaufmann, 1989; Mussel and Schneider, 2019a, 2019b). In the biological case, this membrane interface consists of lipids, proteins, carbohydrates, ions, water, etc. When the state of a membrane is changed, *e.g.* by compression from the fluid towards the gel phase (liquid-expanded to liquid-condensed in terminology of lipid monolayers), the system passes through such a transition and a characteristic maximum of the susceptibilities is crossed (*e.g.* compressibility; see red curve in Fig. 1(a)). Furthermore, the framework predicts a relation between propagation velocity *c* and compressibility $\kappa_S$ of the membrane (see Eq. (1) and for more details: (Griesbauer et al., 2009; Kappler et al., 2017)). As a basis for the analysis herein we use isothermal state diagrams, which are not strictly correct for all types of disturbances of the membrane (slow up to adiabatic state changes), but the transition and its consequences will be present in all cases (Mussel and Schneider, 2019a). To use this theory and to arrive at testable predictions for biology, we start with hypothetical phase diagrams that were inspired from lipid model membranes (Fig. 1).

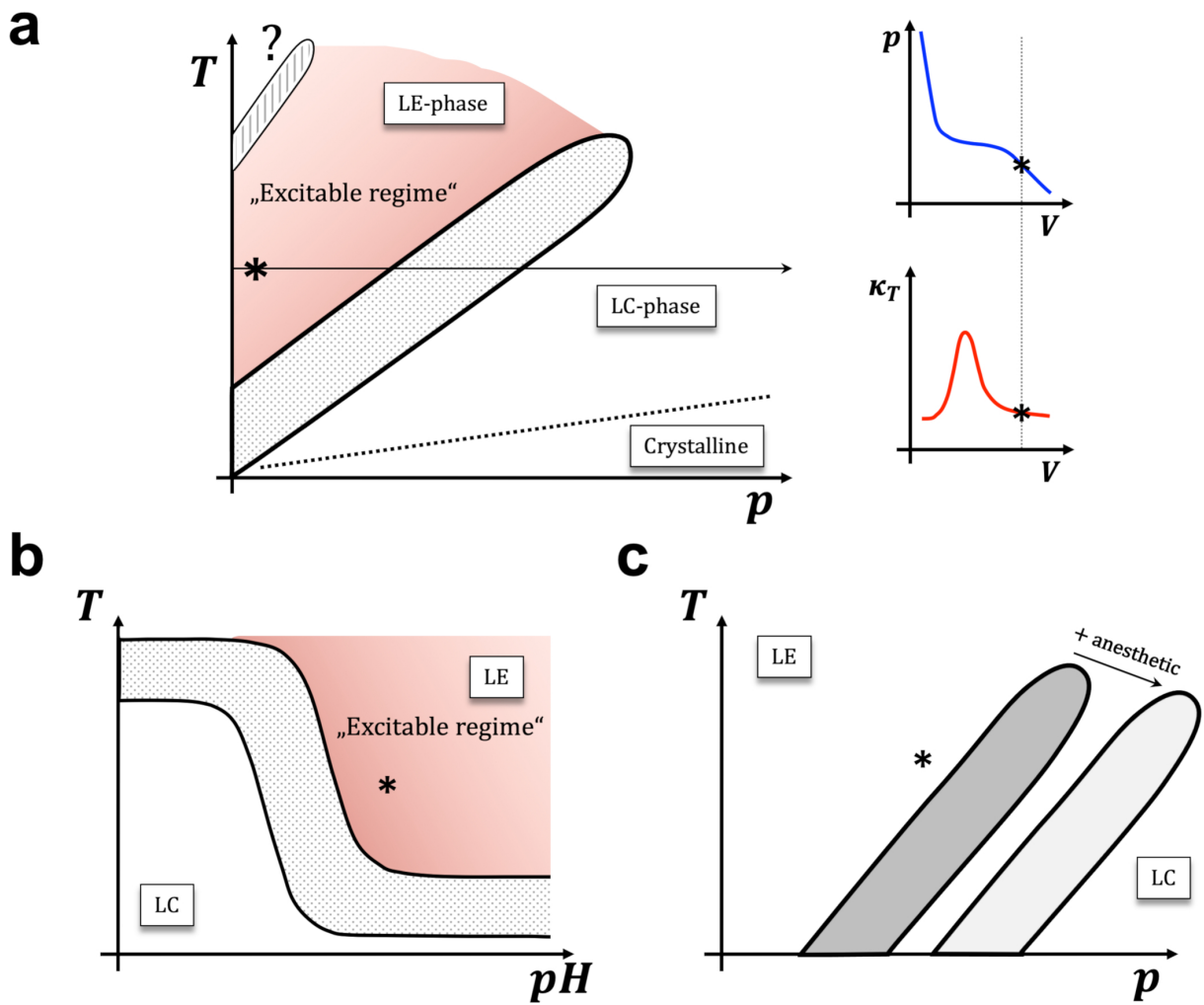

**Figure 1. The relation between state and excitability as summarized in a putative phase diagram. (a)** The cell membrane is excitable if the resting state (asterisk) is in the vicinity of an ordered-disordered transition (*e.g.* from liquid expanded (LE) to liquid condensed phase (LC) in liquid monolayers; or from lamellar fluid to gel in lipid bilayers). This is illustrated by a $p-V$ isotherm and the derived isothermal compressibility $\kappa_T$ on the right (the arrow illustrates the respective slice through the phase plane). State changes move the system state (asterisk) through phase space and hence change the physical properties of the membrane. At low $T$ and/or high $p$ the membrane "freezes" into a crystalline-like state. Note: An additional transition may occur at higher temperatures ("heat block"). **(b)** In the $T-pH$ plane the phase boundary is sigmoidal. The underlying reason for this additional nonlinearity is that the headgroups of membrane molecules have a pK (protonation transition). This results in a nonlinear change of the transition temperature $T_m$ with pH. Thus, acidification at constant $T$ and $p$ can move the resting state into the LC phase. **(c)** According to the melting point depression theory (Heimburg, 2018; Heimburg and Jackson, 2007; Ueda and Yoshida, 1999; Wang et al., 2018)), anesthetics leave the resting state in the disordered phase, but increase its distance to the transition.

It is the goal of this work to test, if these state diagrams capture the essentials of the cellular phenomenology (excitability and pulse velocity).

*3.1.1. Predictions*

The resting state of an excitable membrane should be in vicinity of a transition (see asterisk in Fig. 1). The location of the resting state has been chosen as indicated, because ample evidence suggests that biological membranes reside on the fluid side of a transition (Crowe et al., 1989; Heimburg and Jackson, 2005; Melchior and Steim, 1976; Music et al., 2019; Tablin et al., 1996; Wang et al., 2018). In this regime a suitable perturbation can force the cell membrane into the nonlinear regime (*e.g.* a sudden decrease of $T$ or an increase of $p$, *c.f.* Fig. 1). At the same time, quasi-stationary changes of the cellular environment will lead to new resting states and thus to different excitability and pulse velocity. Since the (mean) compressibility of fluid lipid membranes decreases as temperature increases (Evans and Kwok, 1982; Kang and Schneider, 2020; Steppich et al., 2010) we predict – according to (1) – an increase of $c$ with temperature. This has also been confirmed experimentally for nonlinear waves in lipid monolayers (Kang and Schneider, 2020). Cooling, on the other hand, should lead to a slowing down of pulses. At low $T$, the transition regime into the gel or liquid-condensed state in lipid terminology is crossed. Any stimulus which previously "kicked" the system into this transition and thereby triggered a pulse will now force the system state even further away from the nonlinear behavior. Effectively, the system has become *non-excitable*. Such a crossing of a transition could be the reason for different losses of nonlinear excitability that are observed in experiments ("cold block", "acid block", "3D-pressure block", etc.). Taken together, these arguments also provide a simple, testable explanation for an old pharmacological rule as expressed by one of the originators of the receptor concept, J.N. Langley: *"[…] it is extremely common for a drug to stimulate first and then paralyze […]"* (Langley, 1906). Any substance (or other TD force) that is capable of taking the system towards a transition will be called excitatory (→ stimulation). If the system is continuously exposed to this substance (or TD force), however, a transition takes place and as a result the system will lose its non-linear response to this type of stimulus (→ nerve block, paralysis, desensitization, etc.).

In the framework of the "sound pulse"-description of APs (Heimburg and Jackson, 2005; Mussel and Schneider, 2019a; Shrivastava and Schneider, 2014), the phase diagram implies further testable predictions. Cooling and pressurization, for instance, should have a similar impact on the state and thus excitability and velocity of APs (Fig. 1(a)). As a consequence, it should be possible to compensate the cellular effects of pressurization by heating. Of course, this is only the tip of the iceberg, because the membrane state (and thus excitability and velocity) can also be changed by other thermodynamic forces such as $\mu_{H^+}$, $\mu_{Ca^{++}}$, adsorption of molecules (*e.g.* anesthetics, toxins or proteins; note: the impact of membrane proteins on the state has been likened to an alloy (Sackmann, 1995)), transmembrane potential ($\psi$)). In fact, one should imagine a hypothetical equation of state $p(T, pH, \mu_{ion}, \mu_{protein}, \psi)$, which means that there is not only a relation between $T$ and $p$ (Fig. 1(a)), but also one between $T$ and $pH$ (Fig. 1(b)) or between $T$ and the concentration of anesthetics, etc. We proceed to some concrete predictions: based on the phase diagrams of anionic phospholipids, acidification is expected to reduce the velocity of APs, because it typically brings the system state of a negatively charged interface closer to the transition temperature (Fig. 1(b)). Furthermore, it should be possible to compensate this state and thus velocity change by an increase of $T$. Table 1 summarizes several conjectures that directly emerge from the *state-to-function* approach in combination with the proposed state diagrams in Fig. 1. The remainder of the present work is dedicated to test these predictions in living systems.

**Table 1.** Qualitative predictions based on the phase diagrams (Fig. 1). Increase and decrease of parameter is indicated by upward and downward arrows respectively.

|        | Velocity of AP             | Loss of excitability | Triggering of AP |
|--------|----------------------------|----------------------|------------------|
| $T$    | $c \downarrow$ with $T \downarrow$ | $T \downarrow$       | $T \downarrow$   |
| $p$    | $c \downarrow$ with $p \uparrow$   | $p \uparrow$         | $p \uparrow$     |
| $pH$   | $c \downarrow$ with $pH \downarrow$ | $pH \downarrow$     | $pH \downarrow$  |

We proceed as follows: The TD behavior of excitable living systems will be presented in chapters 3.2 – 3.3 Not only the roles of $T, p$ and $pH$ but also the relation to the melting-point-depression-theory of anesthesia will be discussed and quantitative agreement between our predictions and physiology will be presented. In 3.4 evidence for additional transitions in living systems will be provided and in 3.5 we discuss triggering of APs by different TD forces.

### 3.2. Thermodynamic behavior of action potentials

In the following, the TD phenomenology of the pulse propagation medium of the AP will be established. In all likelihood, the propagation medium is the cell membrane (Terakawa and Nakayama, 1985) which self-assembles in water and contains a multitude of molecules (lipids, proteins, carbohydrates, ions, etc.). We assume that the environmental parameters ($T, p, pH, etc.$) act *directly* on the state of this membrane and thereby modify its excitability as well as the velocity of pulses.

*3.2.1. Loss of excitability upon cooling and heating*

When an excitable plant cell (*Chara* internode) was cooled from the reference state, a progressive slowing down of APs was observed herein (Fig. 2) and by others (M. Beilby and Coster, 1976; Blatt, 1974).

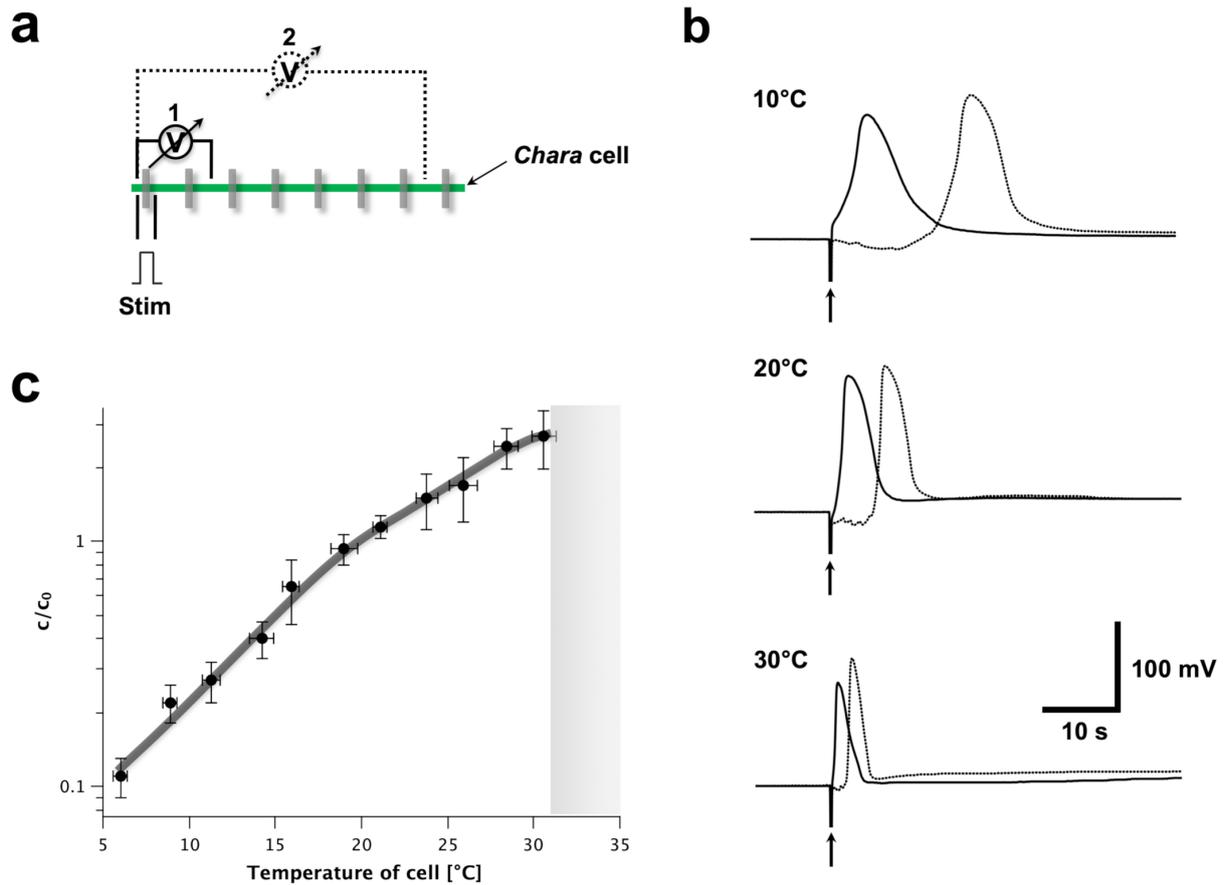

**Figure 2. Temperature dependence of action potential velocity in *Chara*.** **(a)** The pulse velocity was obtained from measurements with two voltage sensors that were fixed along the cell. **(b)** Heating of the cell from 10°C to 30°C increased the pulse velocity. This is evidenced by the decreased delay time of the pulse between two spatially fixed sensors. **(c)** Full temperature dependence. Each data point represents the average ± StDev. of $n = 5 - 27$ measurements in a total of $N = 10$ cells. All data are normalized to the reference propagation velocity $c_0$ ($T = 20°C; pH = 7.0$), which ranged between $4 - 16\ mm\ s^{-1}$ in different cells. The shaded region indicates the regime in which the cell loses excitability ("heat block").

Qualitatively similar results have been reported for axons (Chapman, 1967; Engelhardt, 1951; Franz and Iggo, 1968; Hodgkin and Katz, 1949; Kukita, 1982; Rosenberg, 1978) and muscular tubes (Fillafer and Schneider, 2013). At ~$5°C$, the pulse velocity was reduced to approximately a tenth of that at $20°C$. Upon further cooling, APs often did not propagate along the entire length of the cell any more (*i.e.* the pulse only reached the proximal but not the distal region). This effect was not due to damage sustained to the distal parts, because (*i*) the distal part was still excitable by mechanical/electrical means and (*ii*) propagation over the full cell length was restored upon heating to ~$10°C$. Similar to *Chara*, squid axon preparations are also excitable close to the freezing point of water (Hodgkin and Katz, 1949; Kukita and Yamagishi, 1981). Upon changes of the extracellular monovalent-divalent cation ratio, however, a cold block of excitability becomes apparent above 0°C (Spyropoulos, 1964). In some specimen, this loss of excitability exists under standard conditions (Chapman, 1967). Many other nerves also exhibit such a reversible "cold block" of excitability (Engelhardt, 1951; Franz and Iggo, 1968; Paintal, 1965; Rosenberg, 1978). The block temperature shifts during thermal acclimation (Engelhardt, 1951) and differs between species (Engelhardt, 1951) as well as myelinated and non-myelinated axons (Franz and Iggo, 1968).

Upon heating a *Chara* cell above its growth temperature the velocity increased monotonically up to a maximum at $\sim 30 - 35°C$ (Fig. 2). A further increase of temperature led to cessation of excitability and ultimately cellular death (as evidenced by microscopy). This "heat block" of AP propagation is remarkably universal and has been observed in squid giant axons (Chapman, 1967), *N. ischidiacus* from frog and cat (Engelhardt, 1951), dorsal column of tortoise (Rosenberg, 1978). In squid giant axons (Chapman, 1967; Hodgkin and Katz, 1949) and in blood vessel of blackworms (Fillafer and Schneider, 2013) the heat block was reversible upon cooling. In the present experiments, such reversibility was not observed, presumably due to prolonged exposure ($> 10\ min$) of the cell to the elevated temperatures.

Taken together, the phenomenology of the temperature dependence of $c$ is very well conserved among plant (*Characean*) and animal cells (frog, cat, squid, etc.). The limits of the excitable regime are set by a cold and heat block and the velocity decreases in the vicinity of either block. This is in line with the predictions based on our phase diagram (Fig. 1 and Tab. 1). It is important to note that the cold block has been observed in many living systems that have been studied. In some experiments it was not detected (*e.g.* herein and in some squid axons), presumably, because it is located below the freezing point of the aqueous medium, where experiments are not performed readily.

### 3.2.2. Loss of excitability upon acidification

Alkalization of the extracellular medium of a *Chara* cell up to $pH \sim 10$ had negligible effects on the AP velocity (Fig. 3).

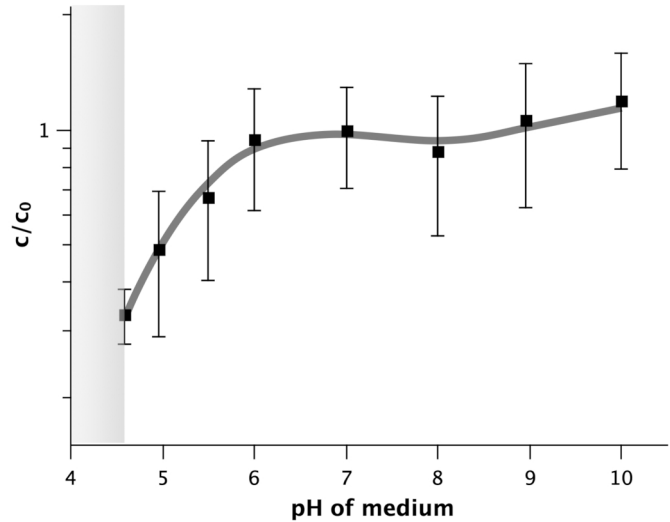

**Figure 3. pH dependence of action potential velocity in *Chara*.** Each data point represents the average ± StDev. of $n = 12 - 17$ measurements in a total of $N = 23$ cells (note: $pH$ 4.6 was an exception with only $n = 3$ data points). All data are normalized to the reference propagation velocity $c_0$ ($T = 20°C; pH = 7.0$). The shaded region indicates loss/block of excitability.

In contrast, when the pH of the extracellular medium was lowered, a progressive slowing down of APs was observed. At $pH \sim 5.0$ the pulse velocity was about half of that at $pH\ 7$. These results are in line with previous reports of a prolongation of APs with decreasing pH in Characean cells (note: a pulse that propagates with a lower velocity will appear prolongated in time) (Beilby, 2007). A decrease in conduction velocity upon acidification has also been observed in cardiac muscle (Vaughan-Williams and Whyte, 1967) and diaphragm (Brody et al., 1991). In *Chara*, further acidification of the extracellular medium ($pH \sim 4.0 - 4.5$) led to a pH-induced block of excitability. Within the timescales of a typical experiment ($10 - 30\ min$) this block was reversible, *i.e.* when the extracellular pH was increased, excitability was reestablished. Such a pH-block has also been observed in other excitable cells, for example, upon acidification by $\sim 1$ pH-unit in heart cells (Harary and Farley, 1963) and squid giant axons (Tasaki et al., 1965). To summarize, the predictions from Tab. 1 (decrease of $c$ with decreasing pH followed by block of excitability) are confirmed.

### 3.2.3. Loss of excitability upon pressurization

To the best of our knowledge the excitability and velocity of the AP in *Chara* has not been studied as a function of the three-dimensional pressure $p$. Since the present analysis, however, can be broadened significantly by this third parameter, we shortly summarize literature data obtained with neurons (for a more extensive review, see *e.g.* (Wann and Macdonald, 1980)). In one of the few works on the subject, Grundfest reported that an increase of hydrostatic pressure first leads to a slight increase of $c$ followed by a slowing down of pulses ($0.55 \cdot c_0$ at $\sim 800\ atm$) (Grundfest, 1936). The decrease of $c$ with an increase of $p$ was confirmed in frog nerves (Ebbecke and Schaefer, 1935; Tasaki and Spyropoulos, 1957) and squid giant axons (decrease by $5 - 15\%$ at ~$340\ atm$) (Spyropoulos, 1957). In the high pressure range ($550 - 1000\ atm$) excitability of frog nerves ceased (Ebbecke and Schaefer, 1935; Grundfest, 1936). This $p$-induced block was reversed upon return to ambient pressure. Taken together, these observations are in line with the predictions set forth in Tab. 1 (decrease of $c$ upon pressurization followed by loss of excitability).

### *3.2.4. The action of anesthetics*

From a medical perspective, it is one of the most intriguing questions how the excitability of cells can be abolished (temporarily). In principle most of our arguments are compatible with the melting point depression theory of anesthesia set forth by Heimburg and others (Heimburg, 2018; Heimburg and Jackson, 2007; Ueda and Yoshida, 1999; Wang et al., 2018). It is a basic premise of this theory that anesthetics preferentially dissolve in the fluid membrane phase and thereby induce a melting point depression. This means that the main transition is shifted to lower temperatures and thus away from the resting state (Fig. 1(c)). Several experimental findings in nerves agree with this assertion. If the *threshold* stimulus is administered, for instance, and subsequently the cell is exposed to anesthetics, the amplitude of the membrane potential pulse decreases until the stimulus-response curve is linear (Tasaki and Spyropoulos, 1957). This transition from a nonlinear to a linear response is in line with the expected removal of the resting state from the nonlinearity. Importantly, however, it should be possible to compensate this increased distance between resting state and depressed melting point by stronger stimulation (Wang et al., 2018) (*i.e.* the nonlinear response of the system is not lost, just shifted). Furthermore, cooling, pressurization and acidification should counteract the effect of anesthetics, because these state changes are capable of moving the resting state towards the lowered melting point ((Heimburg and Jackson, 2007); *c.f.* Fig. 1). If the melting point depression theory is correct, loss of excitability by anesthetics is a different phenomenon as compared to the nerve blocks that have been discussed so far (cold block, acid block, etc.). The latter are a result of crossing the transition and thus cannot be compensated by an increase in stimulus strength (this will only drive the system further into the ordered phase). The physiological outcome of loss of nonlinear excitability can therefore be achieved by at least two different thermodynamic mechanisms.

### *3.2.5. Is nonlinear excitability limited to a phase?*

In Fig. 1, we designated the LE (fluid) phase as the "excitable regime". This was based on the finding that the typical cell membrane at rest is in a fluid state. It was demonstrated, however, that experimental manipulations of a cell – most notably extensive cooling (Chang and Schmidt, 1960) – can lead to stable resting states that are characterized by a depolarized membrane potential level (Chang and Schmidt, 1960; Spyropoulos, 1964; Tasaki, 1959; Terakawa, 1981). It is tempting to speculate that in these experiments the membrane was transferred to an ordered state. If the cellular resting state is indeed below a transition (*c.f.* Fig. 1(a)), one expects that a nonlinear response will now be generated by the opposite stimuli (heating, depressurization, etc.). The underlying reason is the inversion of the asymmetry between resting state and transition. In terms of the electrical stimulus, this prediction is indeed fulfilled: whereas cells exhibit a nonlinear response to outward membrane currents in their typical resting state, the cited studies reported an all-or-none behavior to inward currents (termed "hyperpolarizing response" or "inverted APs" in the literature (Chang and Schmidt, 1960; Spyropoulos, 1964; Tasaki, 1959)). These experiments indicate that the key to understanding the nonlinear response is the transition and its position in relation to the resting state in phase space. Therefore, it must be emphasized that the designation of the LE/fluid phase as the sole "excitable regime" in Fig. 1 is not strictly correct.

### 3.3. Quantitative evidence for the state-to-function hypothesis in excitable systems

Chapter 3.2 has provided the thermodynamic phenomenology of livings systems and demonstrated the agreement with the phase diagram that was borrowed from lipid membrane thermodynamics (Fig. 1). As stated above, the state-to-function hypothesis implies that it is the change in thermodynamic state of the 2D membrane (*not* the parameters and constituents that create it) that makes the difference in biological function. In particular, to be excitable, the membrane must reside near and above a transition in order to be able to access the nonlinearity. In addition to the qualitative predictions (Tab. 1), we now proceed to *quantitative comparisons.* The position of the transition will be given in relation to the cellular resting state. The induction of this transition by $T, p, pH$ will be analyzed and the underlying TD coefficients will be compared with those of lipid model membranes. It will be argued that these coefficients represent a quantitative scale, which can be used to correctly predict the cellular stimulation threshold for $T, p, pH$.

*3.3.1. Location of a transition and thermodynamic relations in living cells*
***TD transition below the growth temperature.*** Several independent investigators have demonstrated that cell membranes of bacteria, plants and animals reside on the fluid (disordered) side of an ordered-disordered transition (Crowe et al., 1989; Heimburg and Jackson, 2005; Melchior and Steim, 1976; Music et al., 2019; Tablin et al., 1996; Wang et al., 2018). These studies have indicated that the midpoint (*e.g.* maximum in $c_p$) of a reversible transition is located $\sim 10 - 30°C$ below the growth temperature of the organism. The latter is in very good agreement with the location of the cold block as obtained from physiological phenomenology ($\sim 15 - 35°C$ below the growth/body temperature of the organism). This suggests that the physical origin of the "cold block" may be the crossing of a membrane transition.
***Shift of membrane transition by p.*** The volume of phospholipid membranes is usually smaller in the ordered phase as compared to the disordered phase (Heimburg, 1998; Winter and Jeworrek, 2009).

Accordingly, an increase in $p$ stabilizes the ordered phase and thereby leads to a rise of melting temperature $T_m$ on the order of $1 - 3°C/100\ atm$ (Tab. 2) (Cossins and Macdonald, 1989; Wann and Macdonald, 1980; Winter and Jeworrek, 2009). This also holds for cellular membranes (Music et al., 2019). From the physiological phenomenology a very similar relation is found. There, an increase of pressure by $+100\ atm$ is equivalent to cooling by $\sim 3°C$ (quantity required for cold block divided by quantity for pressure block, Tab. 2). When pressurizing frog nerves, Grundfest found that an increase of $p$ by $\sim 800\ atm$ leads to cessation of excitability (Grundfest, 1936). This is equivalent to cooling by $\sim 24°C$, which agrees with the quantity required for cold block (cooling by $15 - 35°C$ below growth temperature). Therefore, the cold- and pressure-induced block could have the same underlying cause: the shift of the system state into a transition (compare Fig. 1).

***Shift of membrane transition by pH.*** Variations of pH around a membrane interface will change the degree of protonation of the molecules and thereby will alter the charge state as well as the melting temperature $T_m$ of the membrane. Acidification of typical cellular phospholipids (*e.g.* PS), for instance, is expected to lead to a reduction of membrane charge density and to an increase of $T_m$ that is particularly pronounced in vicinity of the $pK$ (Fig. 1(b)). The same argument applies to any other dissociable molecules (proteins, carbohydrates, etc.) that are part of the membrane interface.

**Table 2.** Parameter changes that lead to loss of excitability in cells and TD coefficients for cells (inferred) and phospholipid membranes (measured).

|  | Loss of excitability in cells | Coefficient in cells | Coefficient in phospholipid membranes |
|---|---|---|---|
| $\|\Delta T\|$ | $15 - 35\ °C$ [a] | | |
| $\|\Delta p\|$ | $800\ atm$ [b] | $\left\|\frac{\Delta T}{\Delta p}\right\| \sim \frac{2 - 4\ °C}{100\ atm}$ | $\left\|\frac{\Delta T_m}{\Delta p}\right\| \sim \frac{1 - 3°C}{100\ atm}$ [d] |
| $\|\Delta pH\|$ | $1 - 3\ units$ [c] | $\left\|\frac{\Delta T}{\Delta pH}\right\| \sim \frac{5 - 35°C}{1\ unit}$ | $\left\|\frac{\Delta T_m}{\Delta pH}\right\| \sim \frac{2 - 20°C}{1\ unit}$ [e] |

[a] (Engelhardt, 1951; Franz and Iggo, 1968; Paintal, 1965; Rosenberg, 1978)
[b] (Grundfest, 1936)
[c] Fig. 2 and (Harary and Farley, 1963; Tasaki et al., 1965)
[d] (Cossins and Macdonald, 1989; Wann and Macdonald, 1980; Winter and Jeworrek, 2009)
[e] (Blume and Eibl, 1979; Heimburg and Jackson, 2007; Träuble, 1977)

From this, it follows that acidification can induce a membrane transition at constant $T$ and $p$ (Cevc, 1991; Träuble, 1977). The typical shift of the main transition of monocomponent phospholipid membranes by $pH$ is $\frac{\Delta T_m}{\Delta pH} \sim \frac{2-20°C}{1\ unit}$ (Blume and Eibl, 1979; Cevc, 1991; Heimburg and Jackson, 2007; Träuble, 1977), which is of the same order of magnitude as the quantity calculated for cells $\left(\frac{\Delta T}{\Delta pH} \sim \frac{5-35°C}{1\ unit}\right)$. Despite the qualitative agreement concerning the sign and order of magnitude of the coefficient, a note of caution should be added. Cellular membranes are a mixture of different lipids and only some of them have headgroups that undergo significant protonation changes in the physiological pH range. For membrane preparations from *E. coli*, for instance, the dependence of $T_m$ on pH $\left(\frac{\Delta T_m}{\Delta pH} \sim \frac{1-2°C}{1\ unit}\right)$ is comparatively small in vicinity of pH 7 (Music et al., 2019). It cannot be excluded that excitable cell membranes represent an ensemble of molecules that has a more pronounced pH dependency. If the *E. coli* response, however, is also representative for excitable cells, there are two possibilities. First, our concept as schematized in Fig. 1 may be incorrect. pH could affect excitability by a different mechanism. This could be proven if an excitable cell membrane has no $T_m$ or if it has a $T_m$, but the latter does not shift with pH as required (Tab. 2). Second, pH may influence the state of a cell membrane by additional ways, which are not captured in the calorimetric experiments, where samples are often sonicated. If pH is only changed on one side of the bilayer, for example, a transmembrane proton gradient will ensue. Since lipid membranes are permeable to protons (Deamer and Bramhall, 1986), this gradient will change the electrical potential across the membrane. Such a change of the electric field will also affect membrane state and $T_m$. A theoretical work predicted that the presence of an electric field should decrease $T_m$ (by $\sim 20°C$ for $\Delta\psi =140$mV), because electrostatic compression stabilizes the fluid phase (Sugar, 1979). In contrast, Antonov et al. demonstrated that $T_m$ of charged and zwitterionic lipid membranes is increased by $5 - 10°C$ for $\Delta\psi \approx 140$mV. These studies have indicated that the effect of an electric field on $T_m$ could be significant in the physiological range of membrane potentials ($\Delta\psi \sim 50 - 200\ mV$). Unfortunately, however, the basis of evidence is small and contradictory.

To summarize, the TD coefficients for lipid membranes and cells $\left(\frac{\Delta T_m}{\Delta p}, \frac{\Delta T_m}{\Delta pH}\right)$, which in principle represent the slopes of phase boundaries in the phase diagrams (Fig. 1), are based on the data of different authors and are given in Tab. 2. The coefficients calculated for cells closely resemble those of artificial phospholipid bilayer membranes. This provides evidence that the two-dimensional membrane state is important for the phenomenology of the biological system. Taken together, these results strongly indicate that cooling, acidification and pressurization take excitable cell membranes towards ($\rightarrow$ reduced velocity) and across a transition ($\rightarrow$ loss of nonlinear excitability). It is a prediction of our approach, that this transition must be observable within the specified parameter ranges.

### *3.3.2. Potential limitations*

It is important to mention limitations of the present approach. First, $\kappa$ of the membrane interface is unlikely to explain the full scale of AP velocity changes (Eq. (1)). When a monocomponent lipid membrane is taken into the main transition, its thermodynamic susceptibilities increase by approximately one order of magnitude as compared to the fluid phase (Albrecht et al., 1978; Evans and Kwok, 1982). According to (1) this sets the scale for the maximal variation of pulse velocity to $\sim 0.3 \cdot c_0$. However, in *Chara* $c$ changes by a factor of $\sim 25$ in the range between 5°C and 30°C (Fig. 2). Similar results have been reported for neurons (Engelhardt, 1951; Franz and Iggo, 1968). In order to account for such large variations, $\kappa$ would have to change by 2-3 orders of magnitude which is physically unrealistic (Kang and Schneider, 2020). This indicates that one or several of the assumptions that have been made (linear sound, no damping, no coupling to bulk phases, etc.) were incorrect. In order to clear up some of these questions, it would be particularly insightful to determine the storage and loss moduli of the cellular interface as a function of $T, pH$ and $p$.

Second, as previously mentioned, we have assumed that $T, pH$ and $p$ act directly on the state of the cell membrane and thereby affect pulse velocity and excitability. This simplification neglects that other materials like the glycocalyx/cell wall or the cytoskeleton experience the same variations of $T, p$ and in a modulated way of $pH$. State changes in these materials could affect the cell membrane by mechanical, chemical or other couplings. In such a coupled system, a membrane response may arise which would be weak/absent if the membrane were studied in isolation. Cellular membranes are intimately associated with cytoskeletal proteins and the latter constitute a significant fraction of the intracellular material. Actin, for instance, accounts for $1 - 10\%$ ($w/w$) of intracellular protein (Lodish et al., 2000), whereas neurofilaments make up 70% of axoplasmic protein (Schmitt and Davison, 1964). Studies with actin (Matthews et al., 2005) and tubulin (Fygenson et al., 1994) have demonstrated that $T$ and $p$ strongly affect the polymerization equilibrium between globular and filamentous form. Depolymerization is favored *in vitro* at low temperature ($< 10°C$ (Fygenson et al., 1994; Matthews et al., 2005)), elevated temperature ($> 25°C$ (Matthews et al., 2005)) and high pressure ($\sim 500 - 1500 \, atm$ (Engelborghs et al., 1976; Garcia et al., 1992)). In line with these reports, disassembly of filaments has been observed in cells upon cooling and pressurization (Begg et al., 1983; Brimijoin et al., 1979; Tilney et al., 1966). Depolymerization occurs in a similar $T-$ and $p-$range as the cold block and $p-$induced block of cellular excitability. The temperature regime in which microtubule disassembly takes place (<15°C (Brimijoin et al., 1979)), for example, coincides with that for the cold block of AP propagation (~10°C (Engelhardt, 1951; Franz and Iggo, 1968). These considerations suggest that a more holistic approach – one that accounts for the coupling between different cellular materials – may be required to fully understand the regulation of cellular excitability. One way to judge if this is indeed necessary for the problem at hand, will be to find out if the blocks of excitability exist in an isolated excitable cell membrane (*e.g.* an axon from which the external material as well as the cytoplasm and cytoskeleton have been removed (Terakawa and Nakayama, 1985)). This could prove that the origin of these phenomena is in the membrane interface.

## 3.4. The heat block of excitability

From the phase diagram in Fig. 1 it is obvious that heating at constant pressure will remove the resting state from the main transition. In the simplest case, compressibility will remain constant or will decrease (velocity is constant or increased respectively). Ultimately, heating should take the system above a critical temperature $T_c$. In this regime, $p - A$ isotherms do not exhibit an inflection point anymore and thus the nonlinearity from the main transition vanishes (Kang and Schneider, 2020). Therefore, nonlinear excitability is not expected at least for stimulation of the system by an increase in $p$ (*c.f.* Fig. 1). However, there exist indications that heating of an excitable cell does not merely take the system further into the disordered phase. It was demonstrated, for example, that the electrical potential and surface tension of a *Nitella* membrane change abruptly at $\sim$35 °$C$ (Ueda et al., 1974). Moreover, the heat capacity profile of *Chara* membranes exhibits a second maximum at $\sim$32 °$C$ (Beljanski et al., 1997), which is in the same range as the heat block temperature found herein (Fig. 2). Fluorescence spectroscopic investigations in crab nerve membrane also indicated a distinct change in emission ratios at $\sim$36°$C$ (Georgescauld et al., 1979). In the vicinity of the heat block in squid axons the electrical membrane capacitance increases drastically (Palti and Adelman, 1969). Such deviations of the susceptibilities ($c_P, C_m$) are characteristic for transitions (Steppich et al., 2010). An increase in susceptibilities should also lead to a slowing down of APs and this was indeed observed in nerve preparations in proximity of the heat block (Chapman, 1967; Rosenberg, 1978). Taken together, these pieces of evidence suggest that an additional transition occurs in the biological membrane in the temperature range of the heat block (Fig. 1(a), indicated by question mark). It will be of fundamental interest to understand the physical mechanism of this phenomenon. Two possibilities should be mentioned. The heat block could be related to protein transitions which are evident from calorimetric investigations of bacterial cells and nervous tissue (onset of denaturation at $\sim$40°$C$ (Music et al., 2019)). Alternatively, the bilayer itself could lose its integrity as the temperature is increased. One of the most prominent examples for breakdown of a lipid membrane is the phenomenon of "electroporation" or "punchthrough" (M. J. Beilby and Coster, 1976). There, the transmembrane electrical potential is increased, *e.g.* by an externally applied current, until the dielectric core is compressed to such an extent that it ruptures. This phenomenon and its induction, however, is not *per se* electrical. Any state change that thins the dielectric core has the potential to induce breakdown (*e.g.* mechanical stretching of the membrane (Needham and Hochmuth, 1989)). In case that the membrane is a closed system, *i.e.* with a constant number of molecules (which can be debated for the case of a cell), heating also leads to membrane thinning by expansion of the area per molecule (Evans and Kwok, 1982; Heimburg, 1998; Steppich et al., 2010). This should favor pore/defect formation, especially since an excitable membrane is already electro-compressed by an intense field (field strength of resting potential $\sim 10^7 \frac{V}{m}$). It was observed indeed that the threshold for electroporation of cells decreases with heating (M. J. Beilby and Coster, 1976; Diaz-Rivera and Rubinsky, 2017). Since electroporation of an excitable membrane also leads to a reversible loss of AP conduction (Gallant and Galbraith, 1997), which is consistent with the phenomenology of the cellular heat block, it would be of particular interest to study if the cellular bilayer breaks down in this temperature range.

### 3.5. Means to stimulate cells ("reception by TD systems")

*3.5.1. Excitation close to a transition*
So far, we considered state changes of the membrane interface in a quasi-stationary manner. The state of a system, however, can also be perturbed rapidly. Such a perturbation will lead to the TD forces being nonzero and this will drive propagation of the disturbance (Kaufmann, 1989). In the vicinity of a transition regime, nonlinear pulses can be induced (Heimburg and Jackson, 2005; Mussel and Schneider, 2019a; Shrivastava et al., 2015; Shrivastava and Schneider, 2014). In the TD approach, this is the usual way how an AP is triggered. In order to be a reasonable stimulus, a perturbation in $T, p, pH, \Psi, etc.$ has to be strong enough to locally induce a transition. Based on the phase diagrams, it is predicted that APs can be triggered by an increase of $p$, a decrease of $T$ and by a decrease of pH because these parameter changes shift the system state into a transition (Fig. 1 and Tab. 2). The closer the resting state of the system is located to the transition, the more "receptive" the interface will appear. Importantly, the membrane state can also be brought into the instable transition regime slowly and this may lead to oscillatory excitations. Furthermore, it has to be emphasized that in a system with mass conservation, *any* local state change will create an asymmetry. Therefore, even parameter changes which move the system away from the transition (decrease of $p$, heating, alkalization) have the potential to induce propagating pulses. These may be linear at first and of opposite polarity, but in principle, could also trigger a nonlinear response. Starting from the resting state in Fig. 1 (a), one example would be rapid local heating, which leads to thermal expansion in the stimulated region (*i.e.* a decrease in density that brings the system state further into the fluid phase). Such an expansion will lead to compression of the non-heated rim. If this compression is of sufficient magnitude it can induce a transition in the rim and thereby induces a nonlinear response.

*3.5.2. Triggering of APs by local cooling*
Whether cells can be excited by cooling was investigated experimentally in *Chara* cells. The external medium of a part of the cell was replaced by the same medium with a lower temperature and triggering of an AP was detected with a voltage sensor at a distance. A sudden drop of temperature of the external medium by $\approx -12°C$ resulted in excitation in 50% of the cases (Fig. 4). This is in line with previous findings in Characean cells ($\Delta T_{thr} \approx -15°C$; (Cook, 1929)). The triggering of APs by local cooling has also been observed in a variety of other excitable cells (Harvey, 1942; Hill, 1935;

Kobatake et al., 1971; Spyropoulos, 1961). It is worth noting that the same decrease in temperature did not trigger an AP if it was applied in a slow manner (*c.f.* Fig. 2). This underlines the difference between a rapid (Fig. 4) and a quasi-stationary (isothermal, Fig. 2) state change.

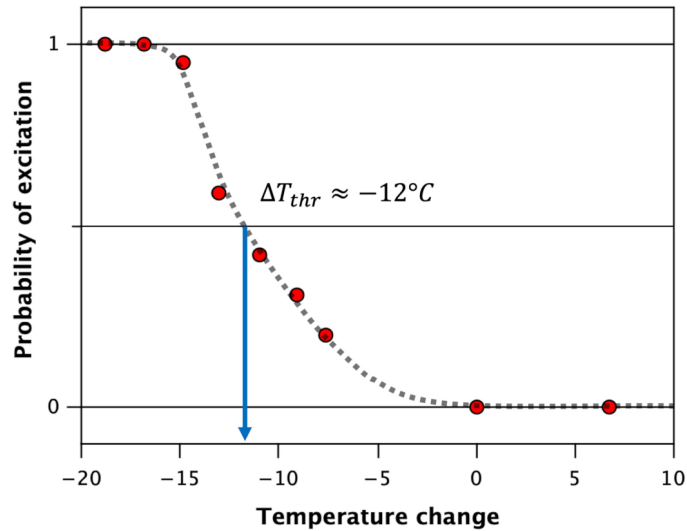

**Figure 4. Triggering of an AP by cooling.** The probabilities of excitation of an AP were calculated based on the results of $n = 117$ stimulation experiments with a total of $N = 9$ cells. The thermal excitation threshold $\Delta T_{thr}$ was defined as the temperature change that triggers an AP in 50% of the cases.

*3.5.3. Excitation by pressurization, acidification and lateral pressure*

In combination with the results in Fig. 4 the TD coefficients (Tab. 2) allow to make predictions about the excitation of APs by non-thermal means. Based on $\frac{\Delta T_m}{\Delta p}$, it can be estimated that a state change similar to a drop of temperature by $\sim 12°C$ (the thermal excitation threshold) can be effected, by pressurization by $\sim 500 - 700\ atm$. Indeed, it has been demonstrated that Characean cells (Harvey, 1942), squid axons (Tasaki and Spyropoulos, 1957) and other neurons (Grundfest, 1936; Wann et al., 1979) can be excited by application of hydrostatic pressure on the order of $350 - 550\ atm$. It has to be noted that in the latter experiments $p$ was increased slowly to avoid heating of the aqueous bulk medium. Thus, the observed oscillatory excitation is probably a consequence of slow transfer of the cellular resting state into the instability of the transition regime.

A similar argument as for $p$ can be applied to excitation by protons. Based on $\frac{\Delta T_m}{\Delta pH}$ (Tab. 1) it can be estimated that a decrease of pH by $2 - 3$ units is required for excitation of a *Chara* cell. This agrees with the experimentally observed excitation threshold (Fillafer and Schneider, 2016). The triggering of APs by acidification is also a very common phenomenon in neurons (Frederickson et al., 1971; Krishtal and Pidoplichko, 1980; Varming, 1999) and it has been proposed that protons could be the excitatory substance in one of the most important synapses, the cholinergic system (Fillafer and Schneider, 2016; Kaufmann, 1980).

Based on the relation between $\tau$ and $p$, it can be predicted furthermore that a sudden increase in lateral membrane pressure by $\sim 5 - 15 \frac{mN}{m}$ should lead to triggering of an AP (see Supporting Information).

## 4. Conclusions and outlook

Herein, we presented evidence that the origin of nonlinear cellular excitability is a TD transition in the membrane interface. Based on experimental data, this transition is located below the growth temperature of the cell and can be induced by cooling, acidification and pressurization. Furthermore, a phase diagram was proposed which allows for an intuitive understanding of several key points (Fig. 1):

*(i)* *the regime of excitable states in a living system,*
*(ii)* *changes of AP velocity,*
*(iii)* *nonlinear stimulation of a cell* (TD parameter change that shifts the state into the transition regime) and
*(iv)* *loss of nonlinear excitability* due to crossing of a transition (cold-block, acid-block, pressure-block) or if the distance between resting state and transition is increased (*e.g.* by melting point depression (Heimburg, 2018; Heimburg and Jackson, 2007; Wang et al., 2018)).

We anticipate that it will be of particular interest to study the state diagrams of biological membranes to identify the mechanisms of evolutionary "fine-tuning" of certain sensitivities of the 2D-interface, *e.g.* to chemicals and light. It must be emphasized as well that the existence of nonlinear pulses is only one consequence of "living close to transitions". In a broader sense, transitions allow for a nonlinear response of all membrane functions (chemical reaction rates, adhesion, permeability, etc.). It follows that the biological system will have a higher response for certain environmental changes, namely the ones that move the system towards a transition. Furthermore, it is likely that transitions play an important role as a boundary for stability of the living system, because they involve drastic changes of the material properties (*e.g.* membrane crystallization). In this regard it is of particular interest that the transitions – and thus the limits of life – are not fixed but dynamic (thermal acclimation/adaptation of membranes (Engelhardt, 1951)). Finally, we consider it perhaps the most intriguing question of all: by which mechanism have cells acquired and retained such a peculiar resting state?

# Acknowledgements

We thank K. Kaufmann for stimulating lectures on the TDs of soft interfaces. His analysis of action potentials in (Kaufmann, 1989) has motivated the present work. The authors also thank I. Foissner for continued support with Characean cells. Moreover, we are grateful for discussions with M. Mussel, S. Fabiunke, K. Kang and S. Shrivastava.

This research did not receive any specific grant from funding agencies in the public, commercial, or not-for-profit sectors.

# References

Aidley, D., 1998. The physiology of excitable cells, 4th ed. Cambridge University Press, Cambridge, UK.

Albrecht, O., Gruler, H., Sackmann, E., others, 1978. Polymorphism of phospholipid monolayers. J. Phys. 39, 301–313.

Begg, D.A., Salmon, E.D., Hyatt, H.A., 1983. The changes in structural organization of actin in the sea urchin egg cortex in response to hydrostatic pressure. J. Cell Biol. 97, 1795–1805. doi:10.1083/jcb.97.6.1795

Beilby, M., Coster, H., 1976. The action potential in Chara corallina: effect of temperature. Aust. J. Plant Physiol. 3, 275.

Beilby, M.J., 2007. Action potential in Charophytes. Int. Rev. Cytol. 257, 43–82.

Beilby, M.J., Coster, H.G.L., 1976. Effect of Temperature on Punchthrough in Electrical Characteristics of the Plasmalemma of Chara corallha.

Beljanski, M. V., Andjus, P.R., Hadži-Pavlović, A., Srejić, R.A., Vučelić, D., 1997. Differential scanning calorimetry of the plasma membrane-enriched fraction in Chara. Plant Sci. 125, 171–176.

Blatt, F., 1974. Temperature dependence of the action potential in Nitella flexilis. Biochim. Biophys. Acta (BBA)-Biomembranes 339, 382–389.

Blume, A., Eibl, H., 1979. The influence of charge on bilayer membranes. Calorimetric investigations of phosphatidic acid bilayers. Biochim. Biophys. Acta 558, 13–21.

Brimijoin, S., Olsen, J., Rosenson, R., 1979. Comparison of the temperature-dependence of rapid axonal transport and microtubules in nerves of the rabbit and bullfrog. J. Physiol. 287, 303–314. doi:10.1113/jphysiol.1979.sp012660

Brody, L.R., Pollock, M.T., Roy, S.H., De Luca, C.J., Celli, B., 1991. pH-induced effects on median frequency and conduction velocity of the myoelectric signal. J. Appl. Physiol. 71, 1878–1885.

Cevc, G., 1991. Isothermal lipid phase transitions. Chem. Phys. Lipids 57, 293–307.

Chang, J.J., Schmidt, R.F., 1960. Prolonged action potentials and regenerative hyperpolarizing responses in Purkinje fibers of mammalian heart. Pflügers Arch. 272, 127–141.

Chapman, R., 1967. Dependence on temperature of the conduction velocity of the action potential of the squid giant axon. Nature 213, 1143–1144.

Cook, S.F., 1929. The effect of sudden changes of temperature on protoplasmic streaming. J. Gen. Physiol. 12, 793–803.

Cossins, A.R., Macdonald, A.G., 1989. The adaptation of biological membranes to temperature and pressure: Fish from the deep and cold. J. Bioenerg. Biomembr. 21, 115–135.

Crowe, J., Hoekstra, F., Crowe, L., Anchordoguy, T., Drobnis, E., 1989. Lipid phase transitions measured in intact cells with Fourier transform infrared spectroscopy. Cryobiology 26, 76–84.

Deamer, D.W., Bramhall, J., 1986. Permeability of lipid bilayers to water and ionic solutes. Chem. Phys. Lipids 40, 167–188.

Diaz-Rivera, R.E., Rubinsky, B., 2017. The effect of temperature on the critical electric field for electroporation: A single cell electroporation device study. J. Eng. Sci. Innov. 1, 83–99.

Ebbecke, U., Schaefer, H., 1935. Über den Einfluß hoher Drucke auf den Aktionsstrom von Muskeln und Nerven. Pflugers Arch. Gesamte Physiol. Menschen Tiere 236, 678–692.

Einstein, A., 1910. Theorie der Opaleszenz von homogenen Flüssigkeiten und Flüssigkeitsgemischen in der Nähe des kritischen Zustandes. Ann. Phys. 338, 1275–-1298.

Engelborghs, Y., Heremans, K.A.H., De Maeyer, L.C.M., Hoebeke, J., 1976. Effect of temperature and pressure on polymerisation equilibrium of neuronal microtubules. Nature 259, 686–689. doi:10.1038/259686a0

Engelhardt, A., 1951. Die Temperaturabhängigkeit der Erregungsleitungsgeschwindigkeit im Kalt- und Warmblüternerven. J. Comp. Physiol. A Neuroethol. 128, 125–128.

Evans, E., Kwok, R., 1982. Mechanical calorimetry of large dimyristoylphosphatidylcholine vesicles in the phase transition region. Biochemistry 21, 4874–9.

Fillafer, C., Paeger, A., Schneider, M.F., 2017. Collision of two action potentials in a single excitable cell. BBA - Gen. Subj. 1861, 3282–3286.

Fillafer, C., Schneider, M., 2013. On the Temperature Behavior of Pulse Propagation and Relaxation in Worms, Nerves and Gels. PLoS One 8, e66773.

Fillafer, C., Schneider, M., 2016. On the excitation of action potentials by protons and its potential implications for cholinergic transmission. Protoplasma 253, 357–365.

Franz, D., Iggo, A., 1968. Conduction failure in myelinated and non-myelinated axons at low temperature. J. Physiol. 199, 319–345.

Frederickson, R.C., Jordan, L.M., Phillis, J.W., 1971. The action of noradrenaline on cortical neurons: Effects of pH. Brain Res. 35, 556–560.

Fygenson, D.K., Braun, E., Libchaber, A., 1994. Phase diagram of microtubules. Phys. Rev. E 50, 1579–1588.

Gallant, P.E., Galbraith, J.A., 1997. Axonal structure and function after axolemmal leakage in the squid giant axon. J. Neurotrauma 14, 811–822.

Garcia, C., Amaral, J., Abrahamson, P., Verjovski-Almeida, S., 1992. Dissociation of F-actin induced by hydrostatic pressure. Eur. J. Biochem. 209, 1005–1011.

Georgescauld, D., Desmazes, J.P., Duclohier, H., 1979. Temperature dependence of the fluorescence of pyrene labeled crab nerve membranes. Mol. Cell. Biochem. 27, 147–153.

Griesbauer, J., Bössinger, S., Wixforth, A., Schneider, M., 2012. Simultaneously propagating voltage and pressure pulses in lipid monolayers of pork brain and synthetic lipids. Phys. Rev. E 86, 061909.

Griesbauer, J., Wixforth, A., Schneider, M.F., 2009. Wave Propagation in Lipid Monolayers. Biophys. J. 97, 2710–2716.

Grundfest, H., 1936. Effects of hydrostatic pressure upon the excitability, the recovery and the potential sequence of frog nerve. Cold Spring Harb. Symp. Quant. Biol. 4, 179–187.

Harary, I., Farley, B., 1963. In vitro studies on single beating rat heart cells. II. Intercellular communication. Exp. Cell Res. 29, 466–474.

Harvey, E.N., 1942. Hydrostatic pressure and temperature in relation to stimulation and cyclosis in Nitella flexilis. J. Gen. Physiol. 25, 855.

Heimburg, T., 1998. Mechanical aspects of membrane thermodynamics. Estimation of the mechanical properties of lipid membranes close to the chain melting transition from calorimetry. Biochim. Biophys. Acta 1415, 147–62.

Heimburg, T., 2018. Phase transitions in biological membranes, in: Demetzos, C. (Ed.), Thermodynamics and Biophysics of Biomedical Nanosystems: Applications and Practical Considerations. Springer-Nature.

Heimburg, T., Jackson, A.D., 2005. On soliton propagation in biomembranes and nerves. Proc. Natl. Acad. Sci. U. S. A. 102, 9790–9795.

Heimburg, T., Jackson, A.D., 2007. The thermodynamics of general anesthesia. Biophys. J. 92, 3159–65.

Hill, S.E., 1935. Stimulation by cold in Nitella. J. Gen. Physiol. 18, 357.

Hill, A. V., 1912. The absence of temperature changes during the transmission of a nervous impulse. J. Physiol. 43, 433–440.

Hodgkin, A., Katz, B., 1949. The effect of temperature on the electrical activity of the giant axon of the squid. J. Physiol. 109, 240–249.

Inoue, I., Kobatake, Y., Tasaki, I., 1973. Excitability, instability and phase transitions in squid axon membrane under internal perfusion with dilute salt solutions. Biochim. Biophys. Acta 307, 471–477.

Kang, K.H., Schneider, M.F., 2020. Nonlinear pulses at the interface and its relation to state and temperature. Eur. Phys. J. E 43, 8.

Kappler, J., Shrivastava, S., Schneider, M.F., Netz, R.R., 2017. Nonlinear fractional waves at elastic interfaces. Phys. Rev. Fluids 2, 114804.

Kaufmann, K., 1980. Acetylcholinesterase und die physikalischen Grundlagen der Nervenerregung, Book. Universität Göttingen, Germany.

Kaufmann, K., 1989. Action potentials and electromechanical coupling in the macroscopic chiral phospholipid bilayer. Caruaru, Brazil.

Kobatake, Y., Tasaki, I., Watanabe, A., 1971. Phase transition in membrane with reference to nerve excitation. Adv. Biophys. 2, 1–31.

Krishtal, O., Pidoplichko, V., 1980. A receptor for protons in the nerve cell membrane. Neuroscience 5, 2325–2327.

Kukita, F., 1982. Properties of sodium and potassium channels of the squid giant axon far below 0°C. J. Membr. Biol. 68, 151–160.

Kukita, F., Yamagishi, S., 1981. Excitation of squid giant axons below 0 degree C. Biophys. J. 35, 243–247.

Langley, J., 1906. On nerve endings and on special excitable substances in cells. Proc. R. Soc. B 78, 170.

Lee, A.G., 1977. Lipid phase transitions and phase diagrams I. Lipid phase transitions. Biochim. Biophys. Acta - Rev. Biomembr. 472, 237–281.

Lodish, H., Berk, A., Zipursky, S., Al., E., 2000. Molecular Cell Biology, 4th ed. W.H. Freeman, New York.

Lühring, H., 2006. Algen unter Strom: Das Experiment. Biol. Unserer Zeit 36, 313–321.

Matthews, J.N.A., Yim, P.B., Jacobs, D.T., Forbes, J.G., Peters, N.D., Greer, S.C., 2005. The polymerization of actin: Extent of polymerization under pressure, volume change of polymerization, and relaxation after temperature jumps. J. Chem. Phys. 123.

Melchior, D., Steim, J., 1976. Thermotropic transitions in biomembranes. Annu. Rev. Biophys. Bioeng. 5, 205–238.

Mosgaard, L.D., Heimburg, T., 2013. Lipid Ion Channels and the Role of Proteins. Acc. Chem. Res. 46, 2966–2976.

Music, T., Tounsi, F., Madsen, S.B., Pollakowski, D., Konrad, M., Heimburg, T., 2019. Melting transitions in biomembranes. BBA - Biomembr. 1861, 183026.

Mussel, M., Schneider, M.F., 2019a. Similarities between action potentials and acoustic pulses in a van der Waals fluid. Sci. Rep. 9, 2467.

Mussel, M., Schneider, M.F., 2019b. It sounds like an action potential: unification of electrical, chemical and mechanical aspects of acoustic pulses in lipids. J. R. Soc. Interface 16, rsif.2018.0743.

Needham, D., Hochmuth, R.M., 1989. Electro-mechanical permeabilization of lipid vesicles. Role of membrane tension and compressibility. Biophys. J. 55, 1001–1009.

Paintal, A., 1965. Block of conduction in mammalian myelinated nerve fibres by low temperatures. J. Physiol. 180, 1–19.

Palti, Y., Adelman, W.J., 1969. Measurement of axonal membrane conductances and capacity by means of a varying potential control voltage clamp. J. Membr. Biol. 1, 431–458.

Rosenberg, M., 1978. Thermal relations of nervous conduction in the tortoise. Comp. Biochem. Physiol. Part A 60, 57–63.

Sackmann, E., 1995. Biological membrane architecture and function, in: Lipowsky, R., Sackmann, E. (Eds.), Structure and Dynamics of Membranes. Elsevier.

Schlaepfer, C.H., Wessel, R., 2015. Excitable Membranes and Action Potentials in Paramecia: An Analysis of the Electrophysiology of Ciliates. J Undergr. Neurosci Educ 14, A82-6.

Schmitt, F.O., Davison, P.F., 1964. Chemical, Structural, and Immunological Studies of Nerve Axon Protein. Berichte der Bunsengesellschaft für Phys. Chemie 68, 887–889. doi:10.1002/bbpc.19640680847

Shrivastava, S., Kang, K., Schneider, M., 2015. Solitary shock waves and adiabatic phase transition in lipid interfaces and nerves. Phys. Rev. E 91, 012715.

Shrivastava, S., Kang, K.H., Schneider, M.F., 2018. Collision and annihilation of nonlinear sound waves and action potentials in interfaces. J. R. Soc. Interface 15.

Shrivastava, S., Schneider, M., 2014. Evidence for two-dimensional solitary sound waves in a lipid controlled interface and its implications for biological signalling. J. R. Soc. Interface 11, 20140098.

Spyropoulos, C.S., 1957. The effects of hydrostatic pressure upon the normal and narcotized nerve fiber. J. Gen. Physiol. 40, 849–857.

Spyropoulos, C.S., 1961. Initiation and abolition of electric response of nerve fiber by thermal and chemical means. Am. J. Physiol. 200, 203–208.

Spyropoulos, C.S., 1964. The role of temperature in the process of excitation of biological membranes. Nuovo Cim. 34, 1837–1839.

Steppich, D., Griesbauer, J., Frommelt, T., Appelt, W., Wixforth, A., Schneider, M., 2010. Thermomechanic-electrical coupling in phospholipid monolayers near the critical point. Phys. Rev. E 81, 061123.

Sugar, I.P., 1979. A theory of the electric field-induced phase transition of phospholipid bilayers. BBA - Biomembr. 556, 72–85.

Tablin, F., Oliver, A.E., Walker, N.J., Crowe, L.M., Crowe, J.H., 1996. Membrane phase transition of intact human platelets: Correlation with cold-induced activation. J. Cell. Physiol. 168, 305–313.

Tasaki, I., 1959. Resting and action potentials of reversed polarity in frog nerve cells. Nature 184, 1574–1575.

Tasaki, I., Singer, I., Takenaka, T., 1965. Effects of internal and external ionic environment on excitability of squid giant axon. A macromolecular approach. J. Gen. Physiol. 48, 1095–123.

Tasaki, I., Spyropoulos, C., 1957. Influence of changes in temperature and pressure on the nerve fiber, in: Johnson, F. (Ed.), Influence of Temperature on Biological Systems. The American Physiological Society, Baltimore (USA), pp. 201–220.

Terakawa, S., 1981. Periodic responses in squid axon membrane exposed intracellularly and extracellularly to solutions containing a single species of salt. J. Membr. Biol. 63, 51–59.

Terakawa, S., Nakayama, T., 1985. Are axoplasmic microtubules necessary for membrane excitation? J. Membr. Biol. 85, 65–77.

Tilney, L.G., Hiramoto, Y., Marsland, D., 1966. Studies on the microtubules in heliozoa. III. A pressure analysis of the role of these structures in the formation and maintenance of the axopodia of Actinosphaerium nucleofilum (Barrett). J. Cell Biol. 29, 77–95.

Träuble, H., 1977. Membrane electrostatics, in: Abrahamsson, S., Pascher, I. (Eds.), Structure of Biological Membranes. Springer, pp. 509–550.

Ueda, I., Yoshida, T., 1999. Hydration of lipid membranes and the action mechanisms of anesthetics and alcohols. Chem. Phys. Lipids 101, 65–79.

Ueda, T., Muratsugu, M., Inoue, I., Kobatake, Y., 1974. Structural changes of excitable membrane formed on the surface of protoplasmic drops isolated from Nitella. J. Membr. Biol. 18, 177–186.

Varming, T., 1999. Proton-gated ion channels in cultured mouse cortical neurons. Neuropharmacology 38, 1875–81.

Vaughan-Williams, E., Whyte, J., 1967. Chemosensitivity of cardiac muscle. J. Physiol. 189, 119–137.

Wang, T., Mužić, T., Jackson, A.D., Heimburg, T., 2018. The free energy of biomembrane and nerve excitation and the role of anesthetics. Biochim. Biophys. Acta - Biomembr. 1860, 2145–2153.

Wann, K., MacDonald, A., Harper, A., 1979. The effects of high hydrostatic pressure on the electrical characteristics of Helix neurons. Comp. Biochem. Physiol. 64A, 149–159.

Wann, K.T., Macdonald, A.G., 1980. The effect of pressure on excitable cells. Comp. Biochem. Physiol. 66A, 1–12.

Wayne, R., 1994. The excitability of plant cells: With a special emphasis on Characean internodal cells. Bot. Rev. 60, 265–367.

Wilke, E., 1912. Das Problem der Reizleitung im Nerven vom Standpunkte der Wellenlehre aus betrachtet. Pflügers Arch. Eur. J. Physiol. 35–38.

Winter, R., Jeworrek, C., 2009. Effect of pressure on membranes. Soft Matter 5, 3157.

Wunderlich, B., Leirer, C., Idzko, A., Keyser, U., Wixforth, A., Myles, V., Heimburg, T., Schneider, M., 2009. Phase-state dependent current fluctuations in pure lipid membranes. Biophys. J. 96, 4592–4597.

## Supporting information

**Predictions about membrane tension**

The 3D pressure that is required to compress a 2D system scales as $P_{3D} \sim P_{2D}/d$ (where $d$ is the thickness of the interface and is in the range of $\sim nm$). From geometrical considerations, one obtains the relation

$$p = \frac{2\tau}{d_0} \cdot \left(1 + \frac{\Delta d/d_0}{\Delta A/A_0}\right)^{-1} \qquad (S1)$$

with $\Delta d/d_0$ and $\Delta A/A_0$ as the relative change in membrane thickness and area respectively (related to the Poisson ratio). At the main transition of dimyristoylphosphatidylcholine (DMPC), $\Delta d/d_0 \approx -0.16$ and $\Delta A/A_0 \approx 0.25$ (Heimburg, 1998). From this it can be estimated that an equivalent change of state induced by $\Delta p = 100\ atm$ can be achieved by an increase in lateral pressure by $\sim 1.8\ \frac{mN}{m}$. Therefore, it can be predicted that lateral pressurization of a biological membrane by $\sim 15\ \frac{mN}{m}$ should lead to a crossing of the transition regime and to a block of AP propagation. In contrast, a decrease in lateral pressure (i.e. an increase in membrane tension) will remove the resting state from the transition regime.